\documentclass[apj]{emulateapj}

\begin{document}

\title{The efficiency of electron acceleration in collisionless shocks
and GRB energetics} \author{David Eichler\altaffilmark{1}} \author{Eli
Waxman\altaffilmark{2}} \altaffiltext{1}{Physics Department,
Ben-Gurion University, Beer-Sheva 84105, Israel;
eichler@bgumail.bgu.ac.il} \altaffiltext{2}{Physics Faculty, Weizmann
Institute, Rehovot 76100, Israel;}

\begin{abstract}

Afterglow observations are commonly used to determine the
parameters of GRB explosions, the energy $E$, surrounding density
$n$, post-shock magnetic field equipartition fraction $\epsilon_B$
and electron equipartition fraction $\epsilon_e$, under the
frequently made assumption that the efficiency of electron
"injection" into relativistic shock acceleration is high, i.e.
that the fraction $f$ of electrons which undergo acceleration is
$f\approx1$. We show that the value of $f$ can not be determined
by current observations, since currently testable model
predictions for a parameter choice
$\{E'=E/f,n'=n/f,\epsilon'_B=f\epsilon_B,\epsilon'_e=f\epsilon_e\}$
are independent of the value of $f$ for $m_e/m_p\le f\le1$.
Current observations imply that the efficiency $f$ is similar for
highly relativistic and for sub relativistic shocks, and plausibly
suggest that $f\sim1$, quite unlike the situation in the Crab
Nebula. However, values $m_e/m_p\le f\ll1$ can not be ruled out,
implying a factor $m_e/m_p$ uncertainty in determination of model
parameters. We show that early, $\le10$~hr, radio afterglow observations,
which will be far more accessible in the SWIFT era, may provide
constraints on $f$. Such observations will therefore provide a
powerful diagnostic of GRB explosions and of the physics of
particle acceleration in collisionless shocks.

\end{abstract}


\section{Introduction}

Synchrotron emission by shock accelerated particles is central to
our understanding of explosive, high energy astrophysical
phenomena, such as supernova remnants, jets from AGN and quasars,
plerionic nebulae, and $\gamma$-ray burst (GRB) afterglows. GRB
afterglows have provided an unprecedented opportunity for
diagnosing the blast wave and attendant shock acceleration,
because their brevity in the observer's time frame and ultrahigh
Lorentz factors allow rapid evolution of the synchrotron spectrum
which can be observed over a wide span of wavelength regimes in
real time.

The afterglow radiation of $\gamma$-ray bursts (GRBs)
\citep[e.g.][]{AG_ex_review} is naturally explained as due to
synchrotron emission of electrons accelerated in relativistic
collisionless shocks driven by the GRB explosion into the medium
surrounding the GRB progenitor \citep[for reviews
see][]{fireballs1,fireballs2,fireballs3}. The energy released in the
explosion leads to the formation of a diverging shock wave, which 
propagates into the ambient plasma. At sufficiently late time (at times
much longer than the burst duration) all the explosion energy is carried
by the shocked ambient plasma (vanishingly small fraction of the energy remains
in the ejecta produced by the explosion). The radiation is believed to be produced by electrons
of the ambient medium, which are accelerated to high energy as they pass through
the diverging shock.

The dynamics of a
spherical shock wave is determined by the explosion energy $E$,
and by the surrounding medium number density $n$. If the initial
GRB outflow is jet-like, an additional parameter, the jet opening
angle $\theta_j$, is required in order to specify the flow. For a
given shock dynamics, the luminosity and spectrum of emitted
radiation are then determined by the fractions $\epsilon_B$ and
$\epsilon_e$ of shock thermal energy carried, respectively, by
magnetic field and electrons, and by the shape of the electron
distribution function. The fraction of explosion energy $E$
converted to thermal energy in the shock is determined by the
hydrodynamics and is of order unity. The electron and
magnetic field energy densities are therefore proportional to $\epsilon_eE$
and $\epsilon_BE$ respectively. The electron distribution function
is commonly assumed to be a power law
of index $p\equiv-{\rm d}\ln n_e/{\rm d}\ln \varepsilon_e$, where $\varepsilon_e$ is
the electron energy, above some minimum energy $\varepsilon_{e0}$ (We use
$\varepsilon$ to denote single particle energy, and $E$ to denote the
total flow energy).

The processes of magnetic field generation and electron injection in
collisionless shocks are not understood from basic principles and
$\epsilon_B$, $\epsilon_e$ and $\varepsilon_{e0}$ cannot at present be
determined theoretically. Rather, they are treated as free parameters
of the model, constrained by observations. It is important to note
here that $\varepsilon_{e0}$ is, in general, an independent parameter of the
model. It is a function not only of $\epsilon_e$, but also of the
fraction $f$ of electrons assumed to be accelerated to beyond
$\varepsilon_{e0}$. It is commonly assumed, however, that $f=1$, in which case
$\varepsilon_{e0}$ is uniquely determined by the other model parameters.

The observed afterglow synchrotron spectra constrain $p$, as is
well known, and the break in the power law decay of afterglow flux
is widely believed \citep{Rhoads99} to establish the opening angle
$\theta_j$ in terms of $E/n$. Under the assumption that {\it all}
the ambient electrons were injected to beyond $\varepsilon_{e0}$, i.e.
$f=1$, the remaining four parameters,
$\{\epsilon_B,\epsilon_e,n,E\}$, are fixed by the four observables
$\nu_m$ (the frequency of maximum intensity), $F_m$ (the intensity
at $\nu_m$), $\nu_{cool}$ (the synchrotron cooling break), and
$\nu_{a}$ (the self-absorption frequency).

The results that have been deduced about $p$, $\epsilon_B$, and
$\epsilon_e$, are consistent both with current knowledge and current
ignorance about shock acceleration: In bursts where $p$ can be
determined accurately \citep[e.g.][]{W97a,galama98,FWK00,Stanek99}
$p=2.2\pm0.1$ is inferred. This value is consistent with the
theoretical value of $p$ derived for test particle acceleration in
relativistic shocks via the first order Fermi mechanism, assuming
isotropic diffusion of particles in momentum space, $p=2.22\pm0.02$
obtained in numerical calculations \citep{RelShock1,Kirk00,RelShock2},
and $p=20/9$ obtained by a more recent analytic analysis
\citep{Keshet04}. This value of $p$ is not consistent with test
particle results for large angle scattering in relativistic shocks,
which produce very hard spectra. It is, however, consistent with the
value expected in the 100 Mev -10 GeV range by non-linear theory for
cosmic ray-mediated shock \citep{EE85,ED02}. Despite the agreement of
the observed and theoretically derived values of $p$, assuming
isotropic diffusion, it should be kept in mind that questions remain
about diffusive shock acceleration, particularly in regard to
relativistic generalization and in regard to electron injection, and
that there are alternative acceleration processes
\citep[e.g.][]{arons,Nishikawa04,Hededal04}.

It is natural to hope that the values of $\epsilon_B$, $\epsilon_e$
are universal since they are determined by the microphysics of the
collisionless shock. The constancy of $p$ and of $\epsilon_e$ among
different bursts is strongly supported by observations. Universal
values of $p$ and $\epsilon_e$, $p\approx2$ and
$\epsilon_e\approx0.1$, typically inferred from most optical
afterglows, are also inferred from the clustering of explosion
energies \citep{Frail01} and from X-ray afterglow
luminosity\footnote{Apparently deviant values of $p$ \citep{CL99,PK02}
are inferred based on light curves, rather than spectra, and are
sensitive to model assumptions (e.g. they depend on the assumed radial
dependence of the ambient medium density).}
\citep{Freedman01,Berger03}. The value of $\epsilon_B$ is less well
constrained by observations. However, in cases where $\epsilon_B$ can
be reliably constrained by multi waveband spectra, values close to
equipartition are inferred \citep[e.g.][]{FWK00}. Such high values for
$\epsilon_B$ and $\epsilon_e$ are remarkable and beg for an
explanation. The magnetic field required for allowing electron
acceleration and emission of synchrotron radiation may conceivably be
produced in the collisionless shock driven by the GRB explosion by
Weibel instabilities or the like \citep[e.g.][]{BE87,
Gruzinov99,Medvedev99}, or it may be that the accelerated particles
mix with the magnetic field of the fireball itself.

No less surprising is the conclusion by \citet{W97b}, that $\varepsilon_{e0}$ is
close to $\gamma m_pc^2$ and that the low frequency radio spectra
imply that there are relatively few electrons in the decade or two
just below $\varepsilon_{e0}$.  Had the electrons been picked up by shock
acceleration at some much lower energy than $\gamma m_pc^2$, the power
law spectrum imparted by the shock acceleration would have extended
down to much lower energies, and only a small minority of them would
have made it to $\gamma m_pc^2$ or higher. In the case of the Crab
nebula, for example, which contains perhaps the best-studied
relativistic shock wave, this is indeed the case: most of the
electrons in the nebula emit in the radio, and probably have Lorentz
factors of order $10^2$, which is many orders of magnitude lower than
$\gamma m_pc^2$ and even about a factor of $10^2$ below $\gamma
m_ec^2$. More will be said about his below. While this paper is not
aimed at explaining this gaping difference between afterglows and the
Crab Nebula, it motivates us to check the assumption that $f=1$ in the
case of the former.

In any case, we are unable to determine from basic principles the
efficiency of electron "injection" to beyond some threshold energy
well beyond $\gamma m_ec^2$. Even when the number of electrons beyond
some injection threshold $\varepsilon_{e0}$ is known, we are unable to determine
theoretically the fraction $f$ of total electrons that these high
energy electrons represent. It is conceivable that a large fraction,
$1-f\sim1$, of the electron population do not participate at the
acceleration process and remain well below $\varepsilon_{e0}$.  This is
discussed in \S~\ref{sec:degeneracy}. In \S~\ref{sec:signatures} we
discuss observational signatures of the existence of such {\it
non-injected} thermal electrons in GRB-induced blast waves. Our main
results and their implications are summarized in
\S~\ref{sec:discussion}. We discuss both the implications to GRB
phenomenology and the implications for the theory of collisionless
shock acceleration, in particular in the context of constraints
imposed by observations on astrophysical systems other than GRBs.

\section{Model prediction degeneracy}
\label{sec:degeneracy}

\begin{figure}[h!]
\plotone{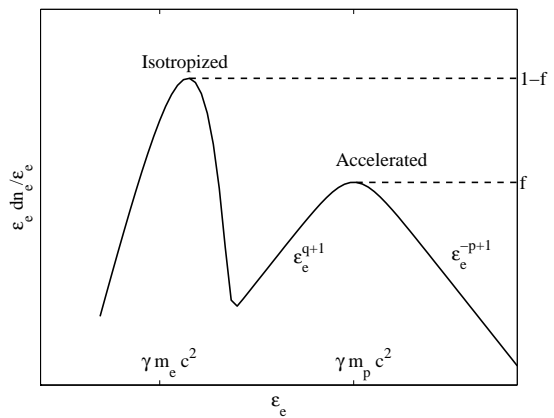}
\caption{A schematic representation of the post shock electron
distribution, for a relativistic shock of Lorentz factor $\gamma$ (or
sub-relativistic shock of velocity $v$). Scattering of electrons
streaming toward the shock with Lorentz factor $\gamma$ (or velocity
$v$) results in post shock "thermal" energy of $\sim\gamma m_ec^2$ (or
$\sim m_e v^2$). A fraction $f$ of the electrons is assumed to be
injected to the acceleration process, which significantly increases
the average energy of these electrons, to $\sim \varepsilon_{e0}$, and produces
a power-law distribution at $\varepsilon_e>\varepsilon_{e0}$. As we show here, afterglow
observations imply $\varepsilon_{e0}\sim\gamma m_pc^2$ in the relativistic phase
and $\varepsilon_{e0}\sim m_pv^2$ in the sub-relativistic phase, but do not
allow to determine $f$. Afterglow observations also require an
electron number density that increases with energy sufficiently fast,
$q\equiv d\ln n_e/d\ln \varepsilon_e>0$, over $\sim1.5$ decades of energy below
$\varepsilon_{e0}$ \citep{W97b}.
\label{fig:dndg}}
\end{figure}

To clarify the issues involved in the electron "injection" problem
let us consider the situation illustrated in
figure~\ref{fig:dndg}, which may arise for a relativistic shock
propagating with Lorentz factor $\gamma\gg1$ (or sub-relativistic
shock propagating with velocity $v\ll c$) into a cold plasma of
protons and electrons (as may be the case for a
shock driven by a GRB explosion into the ISM). 
In the shock frame, a cold stream of
protons and electrons approaches the shock with Lorentz factor
$\gamma$ (velocity $v$). The particles are being scattered at the
shock front, resulting in a velocity distribution which is close
to isotropic behind the shock, thus converting a large fraction of
the kinetic energy of the incoming flow to thermal energy.
Isotropization of the electron and proton incoming flow would lead
to a post shock proton "temperature" $T_p\sim\gamma m_pc^2$ (or
$T_p\sim m_p v^2$), and to a post shock electron "temperature"
$T_e\sim\gamma m_ec^2\ll T_p$ (or $T_e\sim m_e v^2$). In order for
the electrons to gain a significant fraction of the post-shock
thermal energy, some process must couple them to the protons, and
accelerate them to energy $\gg T_e$. This process is yet unknown,
and we can not determine based on theoretical considerations what
fraction of the electrons are being accelerated. Thus, in addition
to $\epsilon_e$, the acceleration process must be described by (at
least) one additional parameter, the fraction $f$ of accelerated
electrons. We show here that afterglow observations imply
$\varepsilon_{e0}\sim\gamma m_pc^2$ in the relativistic phase and
$\varepsilon_{e0}\sim m_pv^2$ in the sub-relativistic phase, but do not
allow  one to determine $f$.

As pointed out in \citet{W97b}, the energy distribution of
electrons below the characteristic acceleration energy, $\varepsilon_{e0}$,
is constrained by radio observations. The slope of the radio
afterglow spectrum observed in several cases \citep[e.g. fig. 1 of
][]{galama98}, $f_\nu\propto\nu^{1/3}$, is consistent with that
expected for radiation emitted by electrons at $\varepsilon_{e0}$ at
frequencies well below their characteristic synchrotron frequency,
which is somewhat below optical at the observation time (typically
of order days). In order for the emission from lower energy
electrons, at $\varepsilon_e<\varepsilon_{e0}$, not to modify this spectrum, $q\equiv
d\ln n_e/d\ln \varepsilon_e>-1/3$ is required. A somewhat more stringent
constraint may be obtained from the requirement that the
self-absorption optical depth produced by these electrons not be
large enough to affect the observed self-absorption frequency
$\nu_a$, $q\equiv d\ln n_e/d\ln \varepsilon_e>2/3$, (provided of course that
$\nu_a$ is unambiguously established by electrons at $\varepsilon_{e0}$).
The uncertainty in the value of $\nu_a$ (and the values of other
characteristic frequencies) determined by observations, relaxes
the latter constraint to $q\equiv d\ln n_e/d\ln \varepsilon_e\gtrsim0$
\citep{W97b}. These statements hold for about 1.5 decades of
energy below $\varepsilon_{e0}$, i.e.  the energy range over which the
electron distribution would affect the radio emission that has
been observed to date. Determining the electron spectrum below
$\varepsilon_{e0}$ is of interest for both GRB phenomenology and particle
acceleration theory, suggesting that more careful analysis of
radio spectra are warranted for obtaining better constraints on
$q$. This is, however, beyond the scope of the current paper,
where we focus on the injection efficiency, $f$.

In what follows we discuss the degeneracy of afterglow model
predictions, showing that the predictions obtained assuming $f=1$ for
some choice of model parameters, $\{E,n,\epsilon_B,\epsilon_e\}$, are
the same as those obtained for any value of $f$, $m_e/m_p\le f\le1$,
and
$\{E'=E/f,n'=n/f,\epsilon'_B=f\epsilon_B,\epsilon'_e=f\epsilon_e\}$. In
\S~\ref{sec:dynamics} we discuss the hydrodynamics of the flow, and in
\S~\ref{sec:radiation} we discuss emission of radiation.

\subsection{Hydrodynamics}
\label{sec:dynamics}

The apparent physical size of the radiation emitting region has been
determined in several cases. During the relativistic stage of shock
expansion, the size of the emitting region has been determined
directly through the observation of the suppression of diffractive
radio scintillation \citep{Goodman97,WKF98}, and through very long
baseline radio interferometry \citep{Taylor04}. At the
sub-relativistic stage, the size was determined indirectly through
modelling the radio spectrum \citep{FWK00,Berger04}. These
observations were used to determine the values of model parameters
that determine the flow pattern, $E$ and $n$, under the assumption of
high efficiency, $f=1$. Modification, due to changes in the value of
$f$, of the values of $E$ and $n$ are therefore allowed provided these
changes do not modify the flow pattern.

We demonstrate here that the velocity field $\vec{v}(\vec{r},t)$
associated with the afterglow stage of the GRB explosion depends on
the explosion energy $E$ and surrounding medium density $n$ only
through the ratio $E/n$. If the surrounding medium density is not
uniform, then for a given functional dependence $g(\vec{r})$ of the
density on coordinates, $n(\vec{r})=n_0g(\vec{r})$ where $n_0$ is some
normalization, then the velocity field $\vec{v}(\vec{r},t)$ depends on
$E$ and $n_0$ only through the ratio $E/n_0$.

It is straightforward to demonstrate the validity of the above
statements through examination of the hydrodynamic equations. These
may be written as
\begin{equation}\label{eq:hydro}
    \partial_\mu T^{\mu\nu}=0,\quad \partial_\mu(nu^{\mu})=0,
\end{equation}
where $u^{\mu}$ is the four-velocity and $T^{\mu\nu}$ is the
energy-momentum tensor. For the GRB afterglow flows, it is appropriate
to assume an ideal fluid flow, i.e. $T^{\mu\nu}=\eta^{\mu\nu}p+(u^\mu
u^\nu/c^2)(p+e)$ where $p$ and $e$ are the fluid pressure and (proper)
energy density respectively, and an ideal gas equation of state,
$e=nm_pc^2+(\hat{\gamma}-1)^{-1}p$ where $\hat{\gamma}$ is the
adiabatic index ($\hat{\gamma}=4/3, 5/3$ for relativistic,
non-relativistic particles respectively). It is now evident that if
$\{u^{\mu}(\vec{r},t),n(\vec{r},t),p(\vec{r},t)\}$ is a solution of
the flow equations, then multiplying the density and pressure by a
constant $K$ and leaving the velocity field unchanged provides another
solution of the equations,
$\{u^{\mu}(\vec{r},t),n'=Kn(\vec{r},t),p'=Kp(\vec{r},t)\}$. Since the
energy-momentum tensor of the new solution $T'^{\mu\nu}$ is related to
the energy-momentum tensor of the original solution $T^{\mu\nu}$ by
$T'^{\mu\nu}=KT^{\mu\nu}$, the energy of the flow in the modified
solution is larger by a factor $K$ compared to the energy of the flow
in the original solution.

The argument given in the previous paragraph proves the statement that
$\vec{v}(\vec{r},t)$ associated with the afterglow stage of the GRB
explosion depends on the explosion energy $E$ and surrounding medium
density $n$ ($n_0$) only through the ratio $E/n$ ($E/n_0$). Since this
argument is, however, rather abstract, it may be useful to examine in
some detail how the afterglow flow is affected at various stages as
$E$ and $n$ are changed. This examination will also be useful for the
discussion of \S~\ref{sec:radiation}. For simplicity, we assume in the
following discussion a uniform density of the surrounding medium (It
is straight forward to generalize the discussion to a non-uniform
density).

Let us first consider the flow associated with a spherical,
relativistic blast wave. When the shock radius $R$ is sufficiently
large, compared to the size of the region of initial energy
deposition, the flow becomes self-similar, with shock lorentz factor
determined by energy conservation, $E\propto\gamma^2 n R^3$ implying
$\gamma\propto(E/n)^{1/2}R^{-3/2}$ \citep{BMK76}. The self-similar
flow is therefore completely determined by the ratio $E/n$.

The early afterglow, on minute time scale, is produced at the onset of
the interaction of relativistic GRB plasma with the surrounding
medium. At this stage, the highly relativistic plasma ejected by the
GRB engine with Lorentz factor $\gamma_i$, the "fireball," drives a
forward shock into the surrounding medium, and a reverse shock is
driven back into the fireball and decelerates it. Once the reverse
shock crosses the fireball plasma shell, the flow approaches the
self-similar behavior described above. This transition stage take
place \citep[e.g.][]{fireballs3} at a radius which is the larger of (i) the
radius at which the self-similar Lorentz factor
$\gamma\propto(E/n)^{1/2}R^{-3/2}$ drops below $\gamma_i$, and (ii)
the radius at which the thickness of the shocked plasma shell in the
self-similar solution, $R/\gamma^2\propto (n/E)R^4$, exceeds the
thickness $\Delta$ of the plasma shell ejected by the GRB. The
transition radius depends, therefore, on $E$ and $n$ only through the
ratio $E/n$.

If the fireball is jet like, rather than spherical, then flow is
well described as a conical section of a spherical fireball as
long as the jet opening angle is $\theta_j>1/\gamma$. In this
case, $E$ should be understood as the "isotropic equivalent
energy", the energy that would have been carried by the blast wave
had it been spherically symmetric. When the fireball decelerates
to $\gamma<1/\theta_j$, the jet is frequently assumed to expand
sideways \citep{Rhoads99}. The condition $\gamma\sim1/\theta_j$
implies that the radius $R_j$ at which sideways expansion begins
is given by $(n/E)R_j^3\propto\theta_j^2$. $R_j$ depends,
therefore, on $E$ and $n$ only through the ratio $E/n$.

During the stage of sideways expansion, the jet does not
significantly propagate radially. Finally, after the stage of
sideways expansion, the flow becomes  sub-relativistic
\citep{FWK00,LnW00}, and, if it becomes a spherical blast wave,
the renewed expansion is described by the Sedov-von Neumann-Taylor
solutions. At this stage the time dependence of the shock radius
is $R\propto (E\theta_j^2/n)^{1/5} t^{2/5}$ (e.g. Chapter XII of
Zel'dovich \& Raizer 2002). Here $t$ is the time and the true
energy of the explosion, corrected for the jet-like geometry, is
$E_T=E\theta_j^2/2$. Thus, at this stage as well the flow depends
on $E$ and $n$ only through the ratio $E/n$.

\subsection{Radiation}
\label{sec:radiation}

Afterglow observations at all stages of flow evolution, from the non
self-similar onset of fireball interaction with surrounding gas,
through the self-similar expansion phase and subsequent jet expansion
phase (if present), and including the final sub-relativistic phase,
are consistent with synchrotron emission of radiation from electrons
accelerated to a distribution of the type shown in fig.~\ref{fig:dndg}
with high efficiency, $f\approx1$. Under the assumption $f=1$, the
values of model parameters, $\{E,n_0,\epsilon_B,\epsilon_e\}$ (as well
as $\theta_j$ and $p$), are determined. Let us now consider what
modifications are introduced by allowing $f\ll1$. We argue that the
emission of radiation from shock accelerated electrons from a flow
with parameter choice
$\{E'=E/f,n'_0=n_0/f,\epsilon'_B=f\epsilon_B,\epsilon'_e=f\epsilon_e,f<1\}$
(and $\theta'_j=\theta_j$, $p'=p$) is similar to that obtained for the
parameter choice $\{E,n_0,\epsilon_B,\epsilon_e\}$ and $f=1$, for any
$f$ in the range $m_e/m_p\le f\le 1$.

Let us first consider the velocity fields of the two flows. The flow
pattern $\vec{v}(\vec{r},t)$ in the modified, $f<1$, flow is similar
to that of the $f=1$ flow, since the energy and density have both been
increased by the same factor $1/f$, leaving the ratio $E/n$ unchanged
(and since $\theta'_j=\theta_j$). Next, we note that the magnetic
field distributions in the two flows are similar. As explained in the
previous section, the energy density in the modified flow is larger
than that in the original flow by a factor $1/f$,
$e'(\vec{r},t)=e(\vec{r},t)/f$. Decreasing the magnetic field
equipartition fraction by a factor $f$, $\epsilon'_B=f\epsilon_B$,
ensures that the magnetic field energy density is similar in both
flows.

Finally, we argue that the density and energy distribution of
accelerated electrons is the same in both flows. The number density of
accelerated electrons is identical in the two flows: The number
density of electrons is larger in the modified flow by a factor $1/f$
compared to that in the original flow, $n'_0=n_0/f$, but only a
fraction $f$ of the electrons in the modified flow are
accelerated. The total energy density in electrons is also the same in
both flows, since $\epsilon'_e e'(\vec{r},t)=\epsilon_e
e(\vec{r},t)$. The fact that the electron energy density and
accelerated electron density is similar in the two flows does not
ensure that the energy distributions of accelerated electrons are
similar, since in the modified flow some part of the electron energy
density is carried by the non shock-accelerated electrons. This part is
small, however, and therefore the energy distributions of the
accelerated electrons are similar in both flows, as long as
$m_e/m_p<f$.

To see this, we note that during the relativistic phase of expansion,
the characteristic Lorentz factor $\gamma_{e0}$ of accelerated
electrons, $\varepsilon_{e0}=\gamma_{e0}m_ec^2$, is approximately determined by
the relation $(1-f)\gamma m_e c^2+f\gamma_{e0}
m_ec^2=\epsilon'_e\gamma m_p c^2=f\epsilon_e \gamma m_p c^2$. Since
afterglow observations imply $\epsilon_e\sim1$ for $f=1$, as long as
$m_e/m_p<f$ we have $\gamma_{e0}\approx\epsilon_e \gamma m_p/m_e$
independent of $f$. During the sub-relativistic regime, $\gamma_{e0}$
is determined by the relation $(1-f)m_e v^2/2+f\gamma_{e0} m_e
c^2=\epsilon'_e m_p v^2/2=f\epsilon_e m_p v^2/2$. Here too,
$\gamma_{e0}$ is independent of $f$, approximately given by
$\gamma_{e0}m_e c^2\approx\epsilon_e m_p v^2/2$, as long as
$m_e/m_p<f$. It is important to note here, that since we have several
examples where afterglow observations cover both relativistic and
sub-relativistic evolution phases \citep[e.g.][]{FWK00,Berger04}, the
efficiency $f$ should be similar at both stages. This independence of
$f$ on $\gamma$ is not necessarily surprising, since $f$ may be, e.g.,
a function of $m_e/m_p$ alone.

Since the two flows have similar velocity fields, similar magnetic
field energy distributions and similar accelerated electron
distributions, the afterglow radiation emitted by the accelerated
electrons is similar for the two flows. The presence of a
non shock-accelerated electron population may, however, modify the radiation
pattern. This issue is discussed in the following section.

\section{Signatures of low efficiency}
\label{sec:signatures}

Consider the possible presence of a large number of "thermal"
electrons that entered the shock at energy $\gamma m_ec^2$ as
measured in the shock frame. Assume that they are heated somewhat
to a typical energy of $\eta \gamma m_e c^2$, where $\eta \ll
m_p/m_e$. Here $\eta$ is a parameter that expresses our ignorance
of the plasma physics that governs the electron heating beyond the
energy $\gamma m_ec^2$, which the electrons bring into the shock
from upstream. The presence of a large population of these
"thermal" electrons at energy $\gamma \eta m_e c^2\ll\gamma
m_pc^2$ (or at $\sim \eta m_e v^2\ll m_pv^2$) may affect the
emitted radiation by producing a new component of emission, or by
producing a large synchrotron self-absorption optical depth, thus
suppressing the emission from accelerated electrons. Since the
energy distribution of the non shock-accelerated electrons does
not extend (by definition) to energies $\gg\gamma \eta m_e c^2$
(or $\gg \eta m_e v^2$), they may affect the emitted radiation at
frequencies $\nu\lesssim\tilde{\nu}_m$, where $\tilde{\nu}_{m}$ is
the characteristic synchrotron emission frequency of electrons of
energy $\eta \gamma m_e c^2$ (or $\eta m_e v^2$). This frequency
is lower by a factor $(\eta m_e/m_p)^2$ than the characteristic
synchrotron emission frequency $\nu_m$ of accelerated electrons at
energy $\gamma m_pc^2$ (or $m_pv^2$, in which case the
self-absorption is a line). Since the time dependence of the
characteristic synchrotron emission frequency of accelerated
electrons typically behaves as $\nu_m\sim10^{18}(t/100{\rm
s})^{-3/2}$, i.e. peaks at X-rays on minute time scale and drops
below the optical on a time scale of 10~hr, the characteristic
synchrotron frequency of the non shock-accelerated electrons drops
from $\sim300 \eta ^2$~GHz on minute time scale to $\sim0.3
\eta^2$~GHz on 3~hr time scale,
\begin{equation}\label{eq:num}
    \tilde{\nu}_{m}\approx1 \eta^2(t/1{\rm hr})^{-3/2}\,{\rm GHz}.
\end{equation}
This implies that the existence of a large population of non
shock-accelerated electrons can be constrained only through early
radio observations, at $\lesssim \eta^{4/3}$~hr delay. Assuming a
spread of several times the thermal energy, so that $\eta \gtrsim
3$, the emission and/or absorption of the thermal electrons could
be caught with radio follow-up observations within several hours
of the GRB. With sufficient preparation, radio follow up observations
may be carried out on minutes time scale (D. Frail, G.
Taylor, private communication).

The specific emissivity $\tilde{j}_m$ of the non
shock-accelerated electrons at $\tilde{\nu}_{m}$ is larger than that of the
accelerated electrons at $\nu_m$ by a factor $1/f$ (since their number
is larger by this factor). Thus, if the optical depth at
$\tilde{\nu}_{m}$ is small, the radio intensity produced by the
non shock-accelerated electrons would be larger by a factor $1/f\gg1$ than
the $\sim1$~mJy peak intensity characteristic of the accelerated
electrons for cosmological GRBs. The synchrotron self-absorption
optical depth at the peak frequency can be estimated using Kirchoff's
law, $\tilde{\tau}_m\propto\tilde{j}_m/\tilde{\nu}_{m}^2\tilde{T}$
where the effective temperature is $\tilde{T}= \eta \gamma m_e
c^2$. From this relation, we find that the ratio of $\tilde{\tau}_m$
to the optical depth $\tau_m$ at $\nu_m$ is (for small $f$)
$\tilde{\tau}_m/\tau_m\approx(m_p/\eta m_e)^5f^{-1}$.  For the
population of accelerated electrons, $\tau_m=(\nu_a/\nu_m)^{5/3}$
where the self-absorption frequency $\nu_a\sim1$~GHz and independent
of time for expansion into uniform medium \citep[e.g.][]{W97b}.
Combining these relations we have, for expansion into uniform medium
and small $f$,
\begin{equation}\label{eq:tau}
    \tilde{\tau}_m\approx \left({m_p\over m_e}\right)^{5/3}\eta^{-5} f^{-1}n_0 (t/1{\rm hr})^{5/2}.
\end{equation}
We have kept here the dependence on the ambient medium number density, $n=10^0n_0{\rm cm}^{-3}$, mainly in order to allow a simple generalization to the case of expansion into a non uniform medium. For expansion into a wind, $n\propto t^{-1}$, and for typical wind parameters $n_0\approx1(t/1{\rm\ day})^{-1}$ \citep{LnW00}. All the results given here can thus be applied to the wind case by using $n_0=1(t/1{\rm\ day})^{-1}$ (note, that eq.~(\ref{eq:num}) is valid for any density, i.e. has no dependence on $n$).

The optical depth at $\tilde{\tau}_m$ is larger than unity for $t>t_a$, where
\begin{equation}\label{eq:t_a}
    t_a\approx 10^{-2}n_0^{-2/5}\eta^{2} f^{2/5}\,{\rm hr}.
\end{equation}
At $t<t_a$, the self-absorption frequency $\tilde{\nu}_a$, where the
optical depth due to the thermal electrons is unity
($\tilde{\tau}_m(\tilde{\nu}_m/\tilde{\nu}_a)^{5/3}=1$), is
\begin{equation}\label{eq:nua}
    \tilde{\nu}_a\approx{m_p\over m_e}n_0^{3/5}\eta^{-1}f^{-3/5}\,{\rm GHz}.
\end{equation}
Finally, the specific intensity at $\tilde{\nu}_m$ is given by
\begin{eqnarray}\label{eq:f_m}
\nonumber
    \tilde{f}_m &\approx& f^{-1}f_m\times\min\left[1,1/\tilde{\tau}_m\right]
    \\&=& f^{-1} \frac{n_0^{1/2}f_m}{1\rm mJy} \times\min\left[1,(t/t_a)^{-5/2}\right]{\rm mJy}.
\end{eqnarray}
Here, $f_m\sim1$~mJy is the peak intensity characteristic of the accelerated
electrons for cosmological GRBs.

The presence of a significant  number of non shock-accelerated
electrons ($f\ll1$) is therefore expected to lead to a large
self-absorption optical depth at frequencies
$\nu\le\tilde{\nu}_m$, strongly suppressing the radio flux at
these frequencies at early time. A sharp rise in the flux at
frequency $\nu$ is expected at $t\approx \eta^{4/3} (\nu/{1\rm
GHz})^{-2/3}$~hr, as $\tilde{\nu}_m$ drops below $\nu$. The ratio of 
fluxes obtained in the presence and in the absence of a non-accelerated
electron population, $\tilde{f}_\nu/f_\nu$, is given for frequencies lower than
the self-absorption frequency of the non-shock accelerated electrons,
$\nu\le\min(\tilde{\nu}_m,\tilde{\nu}_a)$, by the following argument. At 
frequencies where the optical depth is large, the intensity is proportional 
to $\nu^2 T$ where $T$ is the effective electron temperature. For the
non accelerated electrons, $\tilde{T}=\eta\gamma m_e c^2$, and, in the 
absence of electron cooling, $T=\gamma_{e0}m_ec^2$, so that
$\tilde{T}/T\approx\eta m_e/m_p$. At early times, the cooling
time of accelerated electrons is short compared to the dynamical
(expansion) time, and these electrons lose energy and accumulate at lower Lorentz factor,
$\gamma_c$, where the cooling time is comparable to the dynamical time. At this energy
these electrons radiate synchrotron photons at frequency 
$\nu_c=(\gamma_c/\gamma_{e0})^2\nu_m$, and for typical model parameters
\begin{equation}\label{eq:nu_c}
    \frac{\nu_m}{\nu_c}\approx 10 n_0 (t/1{\rm hr})^{-1}.
\end{equation}
Thus, for $\nu\le\min(\tilde{\nu}_m,\tilde{\nu}_a)$ the flux suppression factor
is given by
\begin{equation}\label{eq:f_ratio}
    \frac{\tilde{f}_\nu}{f_\nu}=\frac{\eta m_e}{m_p}
       \max\left[1,\left(\frac{\nu_m}{\nu_c}\right)^{1/2}\right]
       \max\left[1,\left(\frac{\nu}{\nu_a}\right)^{5/3}\right].
\end{equation}
The last term on the rhs accounts for the fact that
$f_\nu\propto\nu^{1/3}$ (rather than $f_\nu\propto\nu^{2}$) for $\nu>\nu_a$.
Cooling of electrons increases $\nu_a$ by a factor $\gamma_{e0}/\gamma_c=(\nu_m/\nu_c)^{1/2}$, 
compared to the case where cooling is unimportant, which implies
\begin{equation}\label{eq:nu_a}
    \nu_a \approx n_0^{3/5}\max\left[1,3n_0^{1/2}(t/{\rm 1 hr})^{-1/2}\right]{\rm GHz}.
\end{equation}

The presences of a large number of non shock-accelerated electrons may
be detected through their radio emission only if this emission takes place
in an optically thin regime, i.e. at frequencies $\tilde{\nu}_a<\nu<\tilde{\nu}_m$ for $\tilde{\nu}_a<\tilde{\nu}_m$. 
Examining eqs.~(\ref{eq:nua}) and~(\ref{eq:num}), we find that $\tilde{\nu}_a<\tilde{\nu}_m$
is possible only for $\eta\gg1$ and moderate $f^{-1}$. For $\eta=10$ and $f=10^{-1}$, for example, the
flux at $\tilde{\nu}_{m}\approx400 (\eta/10)^2(t/0.4{\rm hr})^{-3/2}\,{\rm GHz}$ is $\sim10(10f)^{-1}$~mJy up to $t\approx 0.4 (\eta/10)^{2} (10f)^{2/5}$~hr. For $\tilde{\nu}_a<\nu<\tilde{\nu}_m$ and $\nu>\nu_a$ we have
\begin{equation}\label{eq:f_ratio1}
    \frac{\tilde{f}_\nu}{f_\nu}=\left(\frac{\eta m_e}{m_p}\right)^{-2/3}f^{-1}
       \min\left[1,\left(\frac{\nu_m}{\nu_c}\right)^{-1/3}\right].
\end{equation}

Finally, it is useful to give an estimate of the amplitude of the radio flux typically expected on the relevant time scale. For $\nu>\nu_a$, the flux expected (in the absence of non shock accelerated electrons) is
\begin{eqnarray}\label{eq:f_nu}
\nonumber
    f_\nu&\approx& \max\left[1,\left(\frac{\nu_m}{\nu_c}\right)^{-1/3}\right]
    \left(\frac{\nu}{\nu_c}\right)^{1/3}f_m
\\ \nonumber &\approx& 30 n_0^{5/6}\frac{f_m}{1\rm mJy}\left(\frac{\nu}{10\rm\ GHz}\right)^{1/3}(t/{\rm 1 hr})^{1/6}\,\mu{\rm Jy}
\\ &\times& \max\left[1,\left(\frac{\nu_m}{\nu_c}\right)^{-1/3}\right].
\end{eqnarray}

\section{Discussion}
\label{sec:discussion}

We have shown that current afterglow observations do not allow one to
determine the efficiency of electron acceleration in GRB shocks,
i.e. to determine the fraction $f$ of electrons that are "injected" to
participate in the process of shock acceleration. While afterglow
observations imply that some fraction, $f$, of the electrons'
population is accelerated to a characteristic energy $\varepsilon_{e0}$ comparable to the
post shock proton temperature, $\varepsilon_{e0}\approx\gamma m_p c^2$ for relativistic
shocks of Lorentz factor $\gamma$ or $\varepsilon_{e0}\approx m_p v^2/2$ for non-relativistic
shocks of velocity $v$, a large fraction, $1-f\sim1$, of the electron
population may be "left behind" at low energy comparable to the
kinetic energy of the electrons propagating to the shock, $\gamma m_e
c^2$ for relativistic shocks or $m_e v^2/2$ for non-relativistic
shocks. The resulting electron energy distribution is qualitatively
described in figure~\ref{fig:dndg}. Currently testable afterglow
predictions of a model with parameter choice
$\{E'=E/f,n'_0=n_0/f,\epsilon'_B=f\epsilon_B,\epsilon'_e=f\epsilon_e,f<1\}$
are similar to those obtained for the parameter choice
$\{E,n_0,\epsilon_B,\epsilon_e\}$ and $f=1$, for any $f$ in the range
$m_e/m_p\le f\le 1$. This implies an uncertainty of factor $m_e/m_p$
in the determination of model parameters. Afterglow observations do
not constrain, for example, the values of $E$ and $\epsilon_e$, but
rather the values of $fE$ and $\epsilon_e/f$.  Note, that the value
of $\varepsilon_{e0}$ is independent of $f$ (and equals $\approx\gamma m_p c^2$
or $\approx m_p v^2/2$).

The existence  of non shock-accelerated electrons will strongly
affect the predicted radio emission on short, $\lesssim1
\eta^{4/3}$~hr, time scale. Here, $\eta\gamma m_e c^2$ (or $\eta m_e v^2/2$)  
is the characteristic energy of non shock-accelerated electrons ($\eta\ll m_p/m_e$).
For $f\ll1$, a large self-absorption
optical depth at $\tilde{\nu}_{m}\approx1 \eta^2(t/1{\rm
hr})^{-3/2}\,{\rm GHz}$ [eq.~(\ref{eq:tau})] would lead to strong
suppression of the radio flux at lower frequencies [eq.~(\ref{eq:f_ratio})]. As
$\tilde{\nu}_{m}$ drops below an observed frequency $\nu$, at
$t\approx \eta^{4/3} (\nu/{1\rm GHz})^{-2/3}$~hr, the optical
depth at this frequency drops below unity, and a sharp brightening
is expected. For $\eta\gg1$ and moderate $f^{-1}$, the existence of a large population
of non shock-accelerated electrons may be identified through
their radio emission [see eq.~(\ref{eq:f_ratio1}) and the discussion preceding it].
For $\eta\gg1$, the modification of radio emission due to
the presence of non shock-accelerated electrons will persist over
time scales significantly larger than $1$~hr. The radio signature
of these thermal electrons could test for their presence at levels
that are energetically insignificant by a large margin (even
$f^{-1} \lesssim 1$), and therefore otherwise inconspicuous.

It should be pointed out that afterglow observations already provide
interesting constraints on the efficiency of electron
acceleration. First, they require similar efficiency $f$ for both
relativistic and sub-relativistic shocks, since several examples exist
of afterglow observations covering both relativistic and
sub-relativistic evolution phases \citep[e.g.][]{FWK00,Berger04}. This
independence of $f$ on $\gamma$ is not necessarily surprising, since
$f$ may be, e.g., a function of $m_e/m_p$ alone. Second, afterglow
observations imply that the energy of accelerated electrons is
increased to a characteristic energy similar to the protons post-shock
temperature (with power-law extension to high energies). Finally, the
value of $f$ is limited to $f>m_e/m_p$. Early, $\lesssim1$~hr, radio
observations will provide more stringent constraints on the efficiency
$f$ (and will hence remove the degeneracy in determining GRB model
parameters). As mentioned in \S~\ref{sec:degeneracy}, radio spectra
can also be used to constrain the energy distribution of accelerated
electrons at energies below the characteristic acceleration energy,
$\gamma m_p c^2$ \citep{W97b}, providing further constraints on the
acceleration process.

The total, beaming corrected, energy released in cosmological long
duration GRB explosions is typically inferred, assuming $f=1$, to
be $E_T\sim10^{51.5}$~erg
\citep{FWK00,Freedman01,Frail01,Berger03,Berger04}, with a spread
in estimated values of roughly one order of magnitude. Since
afterglow observations do not constrain $E_T$, but rather $E_T/f$,
the true explosion energies are $E_T\sim f^{-1}10^{51.5}$~erg. For
$f\ll 1$, explosion energies $\gg 10^{51.5}$~erg would naively be
inferred for many GRBs. The association of (at least some) GRBs
with supernovae \citep{Galama98b,Stanek03,Hjorth03,Bloom03}
suggests that the total energy is probably not much more than
$10^{51.5}$~erg, ruling out values of $f\ll1$. Using this argument
to infer a conservative lower limit on $f$, the uncertainties in
determining $E_T/f$ from afterglow observations should be
considered. Uncertainties in determining $E$ may arise from
uncertainties in the determination of the observables
$\{\nu_m,F_m,\nu_{cool},\nu_{sa}\}$. The main uncertainty here is
due to uncertainty in the determination of the self-absorption
frequency, which is not known in many cases and determined in the
best cases to within a factor of $\sim2$, leading to uncertainty
in $E$ of a similar magnitude \citep[since
$E\propto\nu_{sa}^{-5/6}$, e.g.][Note, that the uncertainty in
$n\propto\nu_{sa}^{25/6}$ and $\epsilon_B\propto\nu_{sa}^{-5/2}$
is much larger]{WG99}. Moreover, it should be realized that
afterglow models are highly idealized (e.g. assuming simple
geometry) and various effects which are not taken into account
(e.g. acceleration of pre-shocked plasma by a cosmic-ray
precursor) may lead to systematic errors in estimates of model
parameters. Thus, estimates of the energy should be considered as
order of magnitude estimates. Finally, the energy provided by the
supernova to the GRB jet could be higher than that provided to the
supernovae ejecta, whose energy is limited by neutrino cooling.
Altogether, although $f\sim1$ is suggested by the energy derived
from afterglow observations, $f \ge 1/30$ should be considered a
plausible conservative lower limit.

Is there an {\it a priori} reason to suspect that $f$ should be small?
In the case of plerionic nebulae, such as the Crab, typical
shock-accelerated spectra ($p\sim 2.2$) occur above Lorentz factors of
order $\varepsilon_0 \sim 10^4$, and below that the spectra are much flatter, $
1.3\ge p \ge 1.1$ \citep{Weiler78}. Curiously, the low energy end of these spectra
goes well below $ \gamma m_ec^2$ (where the bulk Lorentz factor of
the pre-shock wind can be estimated knowing the total number of
electrons and the total energy that have been deposited into the
nebulae by the wind). These low energy electrons
do not increase the total energy requirements, but clearly comprise
most of the electrons by number. This raises the question of how most
of the electrons in the nebula can have {\it less} energy than they
had flowing into the shock, and strongly suggests some sort of shock
mediation mechanism that redistributes their energy in the form of a
hard power law.

Suppose the same sort of low energy spectra were obtained below
$\gamma m_pc^2$ in GRB blast waves. We can express these low
energy electron populations as $f^{-1}(\eta) \simeq (\eta
m_e/m_p)^{-p+1}$. Basically $f^{-1}(\eta)$ is the number of low
energy electrons at energy $\eta m_e$ relative to those at $m_p$.
If plerion-like low energy spectra were to obtain in GRB
post-afterglow-shock plasmas, it would give values of $f$ of more
than 1/30 at $\eta \ge 1$ but would  violate the constraint
$d\ln n/d\ln E \ge 0$ that seems to exist at least for some GRB
afterglows.
These current limits extend from $1/30 \le \eta m_e/m_p \le 1$.
Exploring the region $ \eta m_e/m_p \le 1/30$ at 1$ \nu_{\rm GHz}$~GHz
will, by equation (2), require radio follow up observations within
$(10^{3.5}/ \nu_{\rm GHz})^{2/3}$ hr. The signature of a thermal
population of electrons would be a) a  radio  "blackout" due to
the low brightness temperature of the thermal electrons, followed
by b) a pre-brightening, as the emitting area increases, followed
by c) a steep decline, as the emitting frequency of the thermal
electrons passes below the observed frequency. All these stages
precede the expected rise associated with the eventual passage of
$\nu_m$ through the observing frequency.

In summary, early observations of emission and self-absorption in
GRB afterglows can provide a diagnostic of the low energy electron
spectra in GRB afterglow shocks. This might either change our
understanding of them or, at least, close some important loopholes
in afterglow theory.

\acknowledgments The authors thank Y. Lyubarsky and D. Frail for helpful
discussions.  Also acknowledged gratefully is a Center of Excellence
grant from the Israel Science Foundation that encouraged this
collaboration. DE acknowledges support from the Israel-U.S.
Binational Science Foundation and from the Arnow Chair of Theoretical
Astrophysics. EW's research is partially supported by a MINERVA grant.

\end{document}